\documentclass{aa}  
\usepackage{graphicx}
\usepackage{txfonts}
\newcommand{\RNum}[1]{\uppercase\expandafter{\romannumeral #1\relax}}
\newcommand{\Fermitool}{{\fontfamily{qcr}\selectfont FermiTool}}

\newcommand{\isotropic}{{\fontfamily{qcr}\selectfont iso\_P8R3\_SOURCE\_V3\_v1}}
\newcommand{\galactic}{{\fontfamily{qcr}\selectfont gll\_iem\_v07}}
\newcommand{\uvotimsum}{{\fontfamily{qcr}\selectfont uvotimsum}}
\newcommand{\uvotsource}{{\fontfamily{qcr}\selectfont uvotsource}}
\newcommand{\fermipy}{{\fontfamily{qcr}\selectfont Fermipy}}
\newcommand{\srcprob}{{\fontfamily{qcr}\selectfont gtsrcprob}}
\newcommand{\astropy}{{\fontfamily{qcr}\selectfont astropy}}
\newcommand{\jetset}{{\fontfamily{qcr}\selectfont JetSeT}}
\usepackage{multicol}
\begin{document}

   \title{Origin of the broadband emission from the transition blazar \\B2 1308+326}

   \subtitle{}
   \author{Ashwani Pandey\inst{1}
    \and
         Pankaj Kushwaha \inst{2}
   \and
         Paul J. Wiita \inst{3}
        \and
          Raj Prince \inst{1}
         \and
           Bo\.zena Czerny\inst{1}
            \and
         C. S. Stalin \inst{4}          
   }

   \institute{Center for Theoretical Physics, Polish Academy of Sciences, Al.Lotnikov 32/46, PL-02-668 Warsaw, Poland\\
              \email{ashwanitapan@gmail.com}
           \and  
        Department of Physical Sciences, Indian Institute of Science Education and Research (IISER) Mohali, Knowledge City, Sector 81, SAS Nagar, Punjab 140306, India      
         \and
            Department of Physics, The College of New Jersey, 2000 Pennington Road, Ewing, NJ 08628-0718, USA 
            \and
            Indian Institute of Astrophysics, Block II, Koramangala, Bangalore 560034, India 
             }

   \date{Received September 15, 1996; accepted March 16, 1997}

 
  \abstract
   {Transition blazars exhibit a shift from one subclass to the next during different flux states. It is therefore crucial to study them to understand the underlying physics of blazars.}
   {We probe the origin of the multi-wavelength emission from the transition blazar B2 1308+326 using the $\sim$ 14-year long $\gamma-$ray light curve from {\it Fermi} and the quasi-simultaneous data from {\it Swift}.}
   {We used the Bayesian block algorithm to identify epochs of flaring and quiescent flux states and modelled the broadband spectral energy distributions (SEDs) for these epochs. We employed the one-zone leptonic model in which the synchrotron emission causes the low-energy part of the SED and the high-energy part is produced by the inverse-Compton (IC) emission of external seed photons. We also investigated its multi-band variability properties and $\gamma-$ray flux distribution, and the correlation between optical and $\gamma-$ray emissions. }
   {We observed a historically bright flare from B2 1308+326 across the optical to $\gamma-$ray bands in June and July 2022. The highest daily averaged $\gamma-$ray flux was (14.24$\pm$2.36) $\times$ 10$^{-7}$ ph cm$^{-2}$ s$^{-1}$ and was detected on 1 July 2022. For the entire period, the observed variability amplitude was higher at low (optical/UV) energies than at high (X-ray/$\gamma-$ray) energies. The $\gamma-$ray flux distribution was found to be log-normal. The optical and $\gamma-$ray emissions are well correlated with zero time lag. The synchrotron peak frequency changes from $\sim 8 \times$ 10$^{12}$ Hz (in the quiescent state) to $\sim 6 \times$ 10$^{14}$ Hz (in the flaring state), together with a decrease in the Compton dominance (the ratio of IC to the synchrotron peak luminosities), providing a hint that the source transitions from a low-synchrotron peaked blazar (LSP) to an intermediate-synchrotron peaked blazar (ISP). The SEDs for these two states are well fitted by one-zone leptonic models. The parameters in the model fits are essentially consistent between both SEDs, except for the Doppler-beaming factor, which changes from $\sim$15.6 to $\sim$27 during the transition.   }
   {An increase in the Doppler factor might cause both the flare and the transition of B2 1308+326 from an LSP to an ISP blazar. }


   \maketitle
%

\section{Introduction}
Blazars are active galactic nuclei with relativistic jets, whose trajectories are closely aligned to the observer's line of sight \citep{1995PASP..107..803U,2017NatAs...1E.194P}. The main characteristics of blazars are the Doppler-boosted non-thermal emission from the jets, the high-amplitude flux variability over the entire electromagnetic spectrum, and radio-to-X-ray polarizations \citep[e.g.][]{1995ARA&A..33..163W,2022MNRAS.510.1809P,2022MNRAS.517.3236R,2022Natur.611..677L}. Blazars are categorized into flat-spectrum radio quasars (FSRQs; EW\footnote{equivalent width of the emission lines in rest frame}$_{rest}$ $> 5$\AA) and BL Lacertae objects (BLLs; EW$_{rest}$ $< 5$\AA) on the basis of their optical/ultraviolet (UV) emission line properties \citep[e.g.][]{1991ApJS...76..813S, 1996MNRAS.281..425M}. The observed non-thermal broadband spectral energy distributions (SEDs) of blazars show double-hump structures. The peak of the low-energy hump lies in infrared to X-ray energies, and that of the high-energy part ranges from GeV to TeV energies \citep[e.g.][]{1998MNRAS.299..433F}. The low-frequency hump is interpreted as the synchrotron emission produced by relativistic electrons within the jet. On the other hand, two different models, leptonic and hadronic models, have been proposed to explain the high-frequency hump. In the leptonic scenario, the high-energy part of the blazar SED is caused by the inverse-Compton (IC) scattering of either the synchrotron photons (synchrotron-self Compton, SSC; e.g. \cite{1985A&A...146..204G,1996ApJ...461..657B}) or by the external photons (external Compton, EC; e.g. \cite[]{1987ApJ...322..650B,1994ApJ...421..153S}) by the same electrons producing synchrotron emission. In hadronic models, in contrast, the high-energy $\gamma-$ray radiation of a blazar is attributed to hadronic processes such as proton and muon synchrotron emissions \citep[e.g.][]{2001APh....15..121M,2013ApJ...768...54B,2015MNRAS.447...36P}. 

Based on the low-energy component peak ($\nu^p_{syn}$) of their SEDs, blazars are further classified into low-synchrotron peaked (LSPs; $\nu^p_{syn} \leq 10^{14}$ Hz ), intermediate-synchrotron peaked (ISPs; 10$^{14} < \nu^p_{syn} < 10^{15}$ Hz), and high-synchrotron peaked (HSPs; $\nu^p_{syn} \geq 10^{15}$Hz) blazars \citep{2010ApJ...716...30A}. FSRQs belong to the LSP class, and BL Lacs can be LSPs, ISPs, or HSPs based on their $\nu^p_{syn}$. It has been found that the blazar population generally follows an empirical trend, known as the blazar sequence, such that $\nu^p_{syn}$ correlates with the $\gamma-$ray peak frequency and is anticorrelated with the Compton dominance (CD; the ratio of the $\gamma-$ray peak luminosity to the low-energy peak luminosity \citep{1998MNRAS.299..433F,2017MNRAS.469..255G,2022Galax..10...35P}). A possible explanation for the blazar sequence is the difference in the electron cooling efficiency, as proposed by \cite{1998MNRAS.301..451G}. The blazar sequence can also be interpreted as an artefact of the differences in Doppler boosting \citep[e.g.][]{2008A&A...488..867N,2017ApJ...835L..38F}. 

In addition to these conventional classifications, certain blazars showed characteristics of both FSRQs and BL Lacs during their different flux states \citep[e.g.][]{2011MNRAS.414.2674G,2013MNRAS.432L..66G,2014ApJ...797...19R,2021ApJ...913..146M}. Blazars that exhibit a transition from FSRQ to BL Lac or vice versa are known as transition blazars. These blazars can be identified by investigating the shape of their broadband SEDs \citep[e.g.][]{2011MNRAS.414.2674G,2013MNRAS.432L..66G} and/or by estimating the EW of the broad emission lines in their optical/UV spectra \citep[e.g.][]{2014ApJ...797...19R,2021ApJ...913..146M}.

B2 1308$+$326 (OP 313) is a high-redshift (z=0.9980$\pm$0.0005; \cite{2010MNRAS.405.2302H}) blazar. It has been observed several times at different wavelengths because of its variable emissions and uncertain classification \citep[e.g.][]{1993ApJ...410...39G,2000A&A...364...43W,2017A&A...602A..29B}. It was initially classified as a BLL \citep{1991ApJ...374..431S} due to its nearly featureless optical spectra \citep{1979ApJS...41..689W}, high optical polarization, and extreme optical variability \citep{1980ARA&A..18..321A}. However, using very long baseline interferometry (VLBI) polarization images, \cite{1993ApJ...410...39G} observed the polarized flux from the inner part of its jet with a position angle perpendicular to the jet. In addition, they also detected tentative superluminal motion in its VLBI jet. Because these are characteristics of a quasar, they classified B2 1308$+$326 as an FSRQ with unusually weak emission lines. Using their VLBI data together with the optical data of \cite{1991ApJ...374..431S}, they suggested that B2 1308$+$326 might be a gravitationally microlensed quasar. However, \cite{1999ApJ...512...88U} did not detect any spatially extended emission from B2 1308$+$326 in the high-resolution imaging observations carried out with HST WFPC2.

The blazar B2 1308$+$326 was observed simultaneously at X-ray, optical, and radio wavelengths by \cite{2000A&A...364...43W}. They found that it could be a radio-selected BLL based on its optical variability and synchrotron peak power, but because of its high bolometric luminosity, high Doppler factor, and variable line emission, it appears more likely to be an FSRQ. 
They concluded that it may be an intermediate or a transitional blazar or a  gravitationally microlensed quasar (due to excess absorption at X-rays). In the $\gamma-$regime, B2 1308$+$326 was detected by the Large Area Telescope (LAT) on board the {\it Fermi} telescope \citep{2013ApJS..209...34A}. It is designated as an FSRQ in the {\it Fermi}-LAT Fourth Source Catalog (4FGL; \cite {2020ApJS..247...33A}).

In this work, we investigate the physical processes that cause the broadband emission of the blazar B2 1308+326 using optical-to-$\gamma-$ray data for a period of $\sim$ 14 yr. The paper is organized as follows. In Section \ref{sec:data} we discuss the steps we used to reduce the multi-wavelength data. Section \ref{sec:results} presents the results of this study. The details of the SED modelling are given in Section \ref{sec:sed_modeling}. A discussion of our results and our conclusions is given in section \ref{sec:diss_con}. Our findings are summarized in Section \ref{sec:sum}.

\section{Observations and data reduction} \label{sec:data}
\subsection{Fermi-LAT data}\label{subsec:fermi}
  We used the $\gamma-$ray data of B2 1308$+$326 measured with the {\it Fermi}-LAT between  4 August 2008, and 12 December 2022 covering $\sim$ 14 yr of {\it Fermi} operations. We adopted the standard LAT data analysis procedures\footnote{\url{https://fermi.gsfc.nasa.gov/ssc/data/analysis/}} to perform the data analysis. 
The data were reduced using the \Fermitool \ version 2.2.0  and the \fermipy \ version 1.2 \citep{2017ICRC...35..824W}. We selected the Pass 8 Data (P8R3) in the energy range of 0.1$-$500 GeV and considered all the SOURCE class events (evclass=128 and evtype=3) in the region of interest (ROI) of 10$^{\circ}$ $\times$ 10$^{\circ}$ centred on the target source position (RA: 197.619, Dec: 32.3455). We applied the standard cuts (zenith angle, $z_{max}$ $<$ 90$^{\circ}$ and ``(DATA\_QUAL$>$0)\&\&(LAT\_CONFIG==1)'') to obtain the good time intervals. Because at low energy (E$<$100 MeV), the effective area of {\it Fermi}-LAT rapidly decreases and the point-spread function (PSF) increases (e.g. at 100 MeV, the 68\% containment angle of the acceptance-weighted PSF is $\sim$5 degrees\footnote{\url{https://www.slac.stanford.edu/exp/glast/groups/canda/lat_Performance.htm}}) we included all the point sources from the 4FGL within 15$^{\circ}$ of the ROI centre, along with the Galactic (\galactic \ ) and extragalactic isotropic diffuse emission ( \isotropic \ ) components in our initial model file. To optimize the spectral parameters, we performed a binned likelihood analysis with eight energy bins per decade and a spatial binning of 0.1$^{\circ}$ per pixel. 

After initial optimization, we removed sources with test statistics (TS) < 1 and allowed the spectral normalization of sources with TS > 10 to vary in the model file. We also set the spectral shape parameters free to vary for sources within 3$^{\circ}$ of the ROI center. The normalizations of the Galactic and extragalactic diffuse components were left free, together with the spectral index of the Galactic diffuse component. We fitted the ROI again, and when the fit converged successfully, we used the best-fitting model file to generate the $\gamma-$ray light curve and spectra. For the SED analyses, we used a larger ROI of 15 degrees to avoid the systematics.

\subsection{Swift-XRT data}\label{subsec:xrt}
The X-ray light curve and spectra of B2 1308$+$326 were generated using the online {\it Swift}-XRT data products generator  tool\footnote{\url{https://www.swift.ac.uk/user_objects/}}. The details of the data reduction process followed by this online tool can be found in \cite{2007A&A...469..379E,2009MNRAS.397.1177E}.

The 0.3-10 keV X-ray spectra of B2 1308$+$326 were first grouped to a minimum of 20 counts per bin using the GRPPHA task of FTOOLS and were then fitted with an absorbed power-law (PL) model (tbabs*powerlaw) in the XSPEC version 12.12.0. While fitting, we fixed the Galactic hydrogen column density, n$_H$, to 1.22 $\times$ 10$^{20}$ cm$^{-2}$ \citep{2013MNRAS.431..394W}. 

\subsection{Swift-UVOT data}\label{subsec:uvot}
During the Swift monitoring, the UVOT instrument observed the FSRQ B2 1308$+$326 in its three optical ($v$, $b$, and $u$) and three ultraviolet ($uvw1$, $uvm2$, and $uvw2$) filters. We downloaded the UVOT data from the HEASARC Data archive\footnote{\url{https://heasarc.gsfc.nasa.gov/cgi-bin/W3Browse/w3browse.pl}} and performed the data analysis using the HEASoft package version 6.29 and the CALDB version 20211108.

We first summed the multiple observations taken in the same filter over a given epoch using the task \uvotimsum \ and then extracted the source magnitudes using the task \uvotsource \. The source magnitudes were extracted from a circular region of radius 5 arcsec centred on the source, while the background magnitudes were derived from a source-free circular region of radius 20 arcsec. The magnitudes were corrected for the Galactic extinction using the $E(B-V)$ value of 0.0115 taken from \cite{2011ApJ...737..103S} and the extinction laws from \cite{1989ApJ...345..245C}. The reddening-corrected magnitudes were then converted into flux densities using the zero points given in \cite{2011AIPC.1358..373B}.

\subsection{Archival data}\label{subsec:archival}
We also used the publicly available g-band optical data of B2 1308$+$326 from the ASAS-SN\footnote{\url{https://asas-sn.osu.edu/}} (All Sky Automated Survey for Supernovae) data archive \citep{2014ApJ...788...48S,2017PASP..129j4502K} to investigate the correlation between optical and $\gamma-$ray emissions.
\begin{figure*}
\centering
\includegraphics[width=18cm, height=8cm]{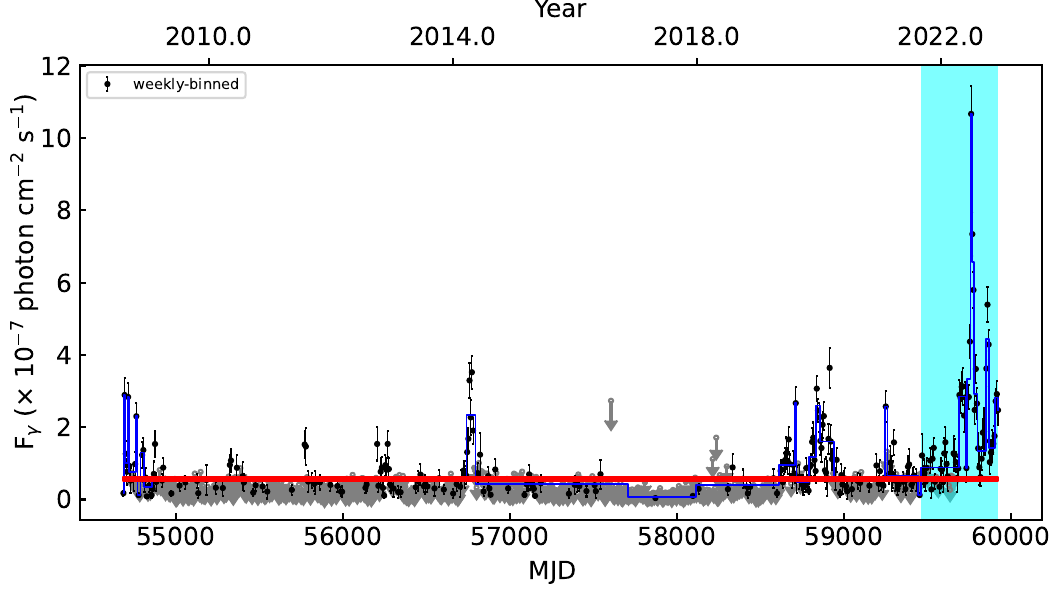} 
\caption{\label{fig:weekly_lc} Weekly averaged $\gamma-$ray light curves of B2 1308$+$326 covering a period of $\sim$ 14 yr. The horizontal red band gives the average flux (5.69 $\times$ 10$^{-8}$ ph cm$^{-2}$ s$^{-1}$) with 1 $\sigma$ uncertainty, and the solid blue line delineates the Bayesian blocks. The shaded region represents the outburst phase.}
\end{figure*}

\begin{figure}
\centering
\includegraphics[width=8.5cm, height=6cm]{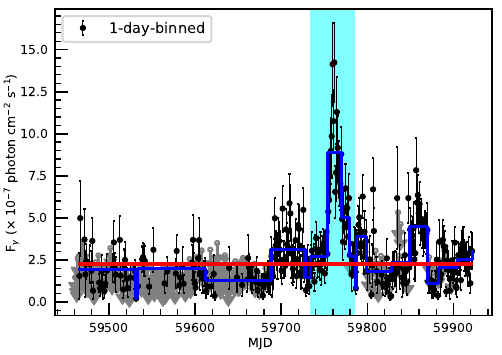} 
\caption{\label{fig:daily_lc}Daily binned $\gamma-$ray light curve for the outburst phase. The shaded region denotes the flaring period. }
\end{figure}

\begin{figure*}
\centering
\includegraphics[width=18cm, height=20cm]{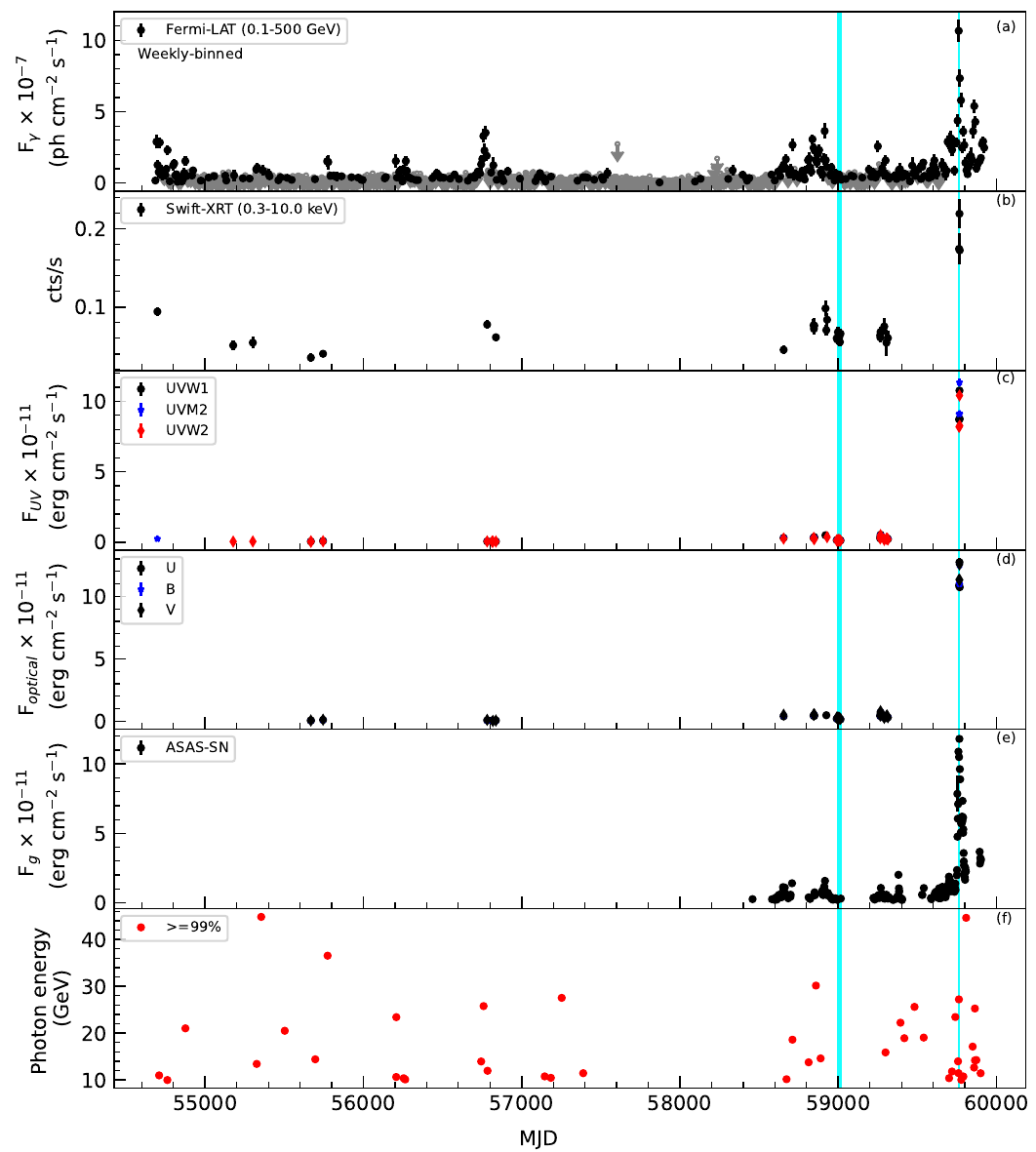}
\caption{\label{fig:multi_lc}Multi-wavelength light curves of B2 1308$+$326 and (bottom panel) the highest photon energies detected at different times. The cyan shaded regions represent the quiescent and flaring periods used for our SED modelling.}
\end{figure*}

\section{Results}\label{sec:results}
\subsection{Flare selection}
The weekly binned $\gamma-$ray light curve of B2 1308$+$326 is shown in Figure \ref{fig:weekly_lc}. We show upper limits at the 95\% confidence limit for the time bins within which either the source is detected with TS $<$ 10 or the flux is lower than or equal to its error $F_i \leq \sigma_i$. The average $\gamma-$ray flux $\overline{F} = (0.57\pm0.05) \times$ 10$^{-7}$ ph cm$^{-2}$ s$^{-1}$,  
estimated using the maximum likelihood analysis over the entire $\sim$ 14 yr monitoring, is shown as a horizontal red band. The source exhibits multiple episodes of high $\gamma-$ray activity, with the highest weekly averaged $\gamma-$ray flux of 10.67 $\times 10^{-7}$ ph cm$^{-2}$ s$^{-1}$ recorded recently on MJD 59761.

There is no generally accepted way of defining a flare in a light curve. However, the Bayesian blocks (BB) algorithm \citep{2013ApJ...764..167S} has recently been used by several authors to delineate flares in blazar light curves \citep[e.g.][]{2019ApJ...877...39M,2022MNRAS.517.2757S}. The algorithm provides a block-wise constant representation of a light curve by identifying the statistically significant variations. We adopted the \astropy\footnote{\url{https://docs.astropy.org/en/stable/api/astropy.stats.bayesian_blocks.html}} implementation of the BB algorithm with a false-alarm probability $p_0 = 0.05$ to identify the flares in the $\gamma-$ray light curves of B2 1308$+$326. Because the BB algorithm can be applied to non-uniform light curves, we ignored the upper limits while applying the BB method to the light curves. The BB representation of the weekly binned $\gamma-$ray light curve of B2 1308$+$326 is shown as solid blue lines in Figure \ref{fig:weekly_lc}, where several flares exceeding the mean flux level are easily recognized. To estimate the duration of the flares and to treat the overlapping flares, we used the HOP algorithm \citep{1998ApJ...498..137E,2019ApJ...877...39M}. The algorithm includes two steps: (1) It identifies a block with a value $F_{BB}$ higher than both the preceding and following blocks as a peak, and (2) it moves downward in both directions from the peak and includes blocks with a value $F_{BB} \geq \Bar{F}$ into the peak to form a HOP group. Using the algorithm, we iteratively searched the $\gamma-$ray light curve for HOP groups and arbitrarily selected only those whose peak value $F_{BB} \geq 5 \times \bar{F}$ for the weekly binned light curve \citep[e.g.][]{2019ApJ...877...39M}. In this way, we found only one HOP group (outburst phase), which spans MJD 59460.16 to 59922.05 ($\sim$ 462 days). For this duration, we reoptimized the spectral model following the steps described in Section \ref{subsec:fermi} and extracted the one-day-binned light curve that is shown in Figure \ref{fig:daily_lc}. We searched the one-day-binned light curve using the BB and HOP algorithm for the flare identification and found one flare that satisfied the arbitrary condition $F_{BB} \geq 2 \times \bar{F}$, which covers the period from MJD 59733.66 to MJD 59786.16 ($\sim$ 52.5 days). During the flare, the highest daily averaged $\gamma-$ray flux of  (14.24$\pm$2.36) $\times$ 10$^{-7}$ ph cm$^{-2}$ s$^{-1}$ was observed on MJD 59761.65 (1 July 2022).

\subsection{Multi-band flux variability}
The $\gamma-$ray to optical energy band light curves of B2 1308$+$326 are plotted in Figure \ref{fig:multi_lc}. The figure shows that B2 1308$+$326 exhibits strong flux variations at all these different energy bands. To quantify these flux variations, we estimated the fractional variability amplitude \citep[e.g.][and references therein]{2003MNRAS.345.1271V,2017ApJ...841..123P}, which is defined as follows:
\begin{equation}
F_{var} = \sqrt{\frac{S^2 - \overline{\sigma_{err}^2}}{{\bar{x}^2}}}.
\end{equation}
The uncertainty in $F_{var}$ can be determined as 
\begin{equation}
err(F_{var}) =  \sqrt{\left( \sqrt{\frac{1}{2N}}\frac{\overline{\sigma_{err}^2}}{\bar{x}^2 F_{var}} \right)  ^ 2+ \left(  \sqrt{\frac{\overline{\sigma_{err}^2}}{N}} \frac{1}{\bar{x}}\right) ^2 },
\end{equation}
where $S^2$ is the sample variance of the light curve, $\bar{x}$ is the mean flux, and  $\overline{\sigma_{err}^2}$ is the mean square~error. The values of $F_{var}$ together with its uncertainty for the different energy band light curves are given in Table \ref{tab:f_var}. At low energies (optical/UV), the blazar B2 1308$+$326 showed more flux variations than at high energies (X-ray/$\gamma-$ray).

We also calculated the shortest flux doubling/halving timescales during the flaring period, as follows:
\begin{equation}\label{eq:doubling_time}
    F(t_{2}) = F(t_{1})\times2^{\Delta t/\tau},
\end{equation}
where F($t_{1}$) and F($t_{2}$) denote the fluxes at times $t_{1}$ and $t_{2}$, respectively, $\Delta t = t_{2}-t_{1}$, and $\tau$ represents the flux doubling or halving timescale. The shortest $\gamma-$ray flux halving timescale is found to be (140.86$\pm$41.15) hr between MJD 59761.66 and 59770.66, while at the optical ($g$) band, we found a shorter flux doubling timescale of (80.71$\pm$2.53) hr between MJD 59757.25 and 59759.32.

\begin{table}
\centering
\caption{\label{tab:f_var}Fractional variability amplitudes (in percent) at different energy bands with a one-day binning. The value of F$_{var}$ for the ASAS-SN g band is estimated for the period of MJD 58458.65-59901.49. For the same duration, the F$_{var}$ for $\gamma-$rays is 67.42$\pm$2.31\%.}
\begin{tabular} {lc} \hline
Band &  F$_{var} (\%)$ \\ \hline
Gamma-ray & 66.03$\pm$ 1.94 \\
Swift-XRT & 51.57$\pm$ 2.15 \\ 
Swift-UVOT u band & 244.56$\pm$ 1.26 \\ 
Swift-UVOT b band & 235.44$\pm$ 1.18 \\ 
Swift-UVOT v band & 226.42$\pm$ 1.35 \\ 
Swift-UVOT m1 band & 240.21$\pm$ 1.19 \\ 
Swift-UVOT w1 band & 242.39$\pm$ 1.12 \\ 
Swift-UVOT w2 band & 260.91$\pm$ 1.09 \\ 
ASAS-SN g band &  144.33$\pm$ 0.55 \\  
\hline
\end{tabular}
\end{table}

\subsection{Highest-energy photon}
We also extracted the arrival time, energy, and probability of the highest-energy photons coming from the source using the tool \srcprob \ on the ULTRACLEAN (evclass=512) event class and a 0.5$^{\circ}$ ROI. We plot the energies of the photons with a probability $>$99\% against their arrival times in the bottom panel of Figure \ref{fig:multi_lc}. Two photons with energies of 44.84 GeV and 44.64 GeV were observed on MJD 55354.54 and MJD 59807.89, respectively. During the 2022 flare, we detected a maximum photon energy of $\sim$ 27 GeV on MJD 59762.30. 

\subsection{Gamma-ray flux distribution}
We investigated the distribution of weekly averaged $\gamma-$ray fluxes of B2 1308$+$326 for bins with TS $\geq$10 and $F_i > \sigma_i$. In this way, we neglected the upper limits of the $\gamma-$ray flux, which are mostly below 0.5 $\times 10^{-7}$ ph cm$^{-2}$ s$^{-1}$ and may form a disjoint population in the distribution. We obtained upper limits instead of a flux value in $\sim$ 64\% time bins for the weekly binned light curve. We first performed the Anderson-Darling (AD) test. The null hypothesis for the AD test is that the sample follows a normal distribution. We applied the AD test on the flux and log\footnote{logarithmic with base 10}-flux distributions and estimated the $p-$values  for the flux and log-flux distributions, which are 1.3e-51 and 0.16, respectively. A $p-$value < 0.01 indicates a clear deviation from a nominally normal distribution. Therefore, the $\gamma-$ray flux distribution of B2 1308$+$326 does not follow a normal distribution on a linear scale. However, on a logarithmic scale, the flux distribution is consistent with normal, implying that the $\gamma-$ray flux distribution of B2 1308$+$326 is essentially log-normal. 

\begin{figure*}
\centering
\includegraphics[width=8cm, height=6.5cm]{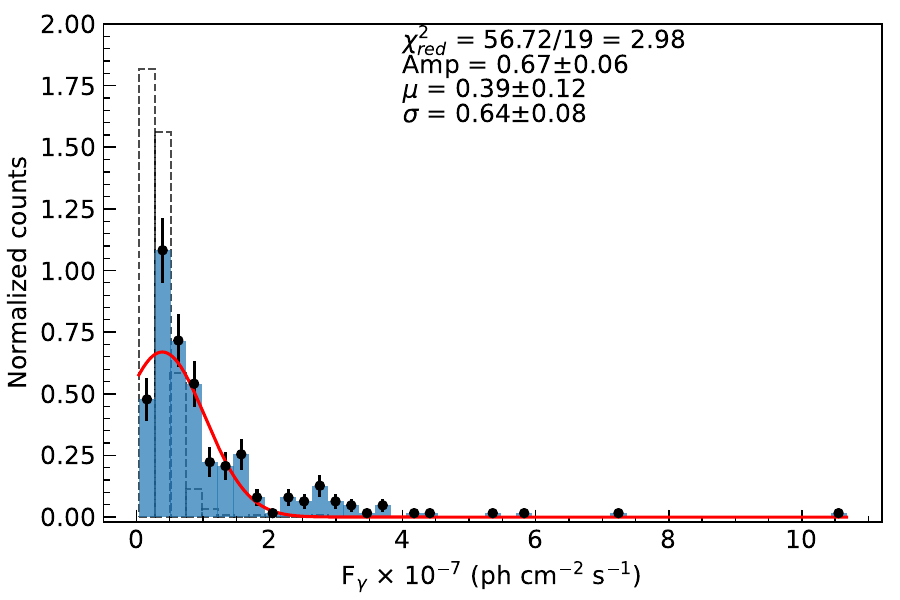}\hfill \includegraphics[width=8cm, height=6.5cm]{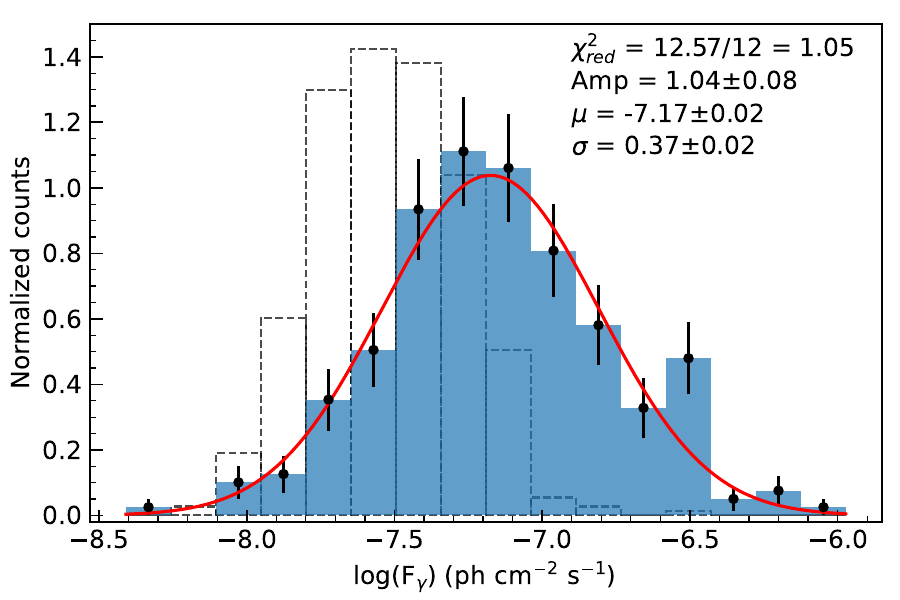}
\caption{\label{fig:flx_dis}Gamma-ray flux distribution of B2 1308$+$326 for the entire $\sim$ 14 yr monitoring period. Left panel: Flux distribution fitted with a Gaussian function on a linear scale (normal distribution). Right panel: Flux distribution fitted with a Gaussian function on a logarithmic scale (log-normal distribution). The results of the fits are mentioned in each plot. The distribution of the upper limits is shown as dashed bars.}
\end{figure*}

We also constructed the normalized histograms of flux values on linear as well as logarithmic scales, shown in Figure \ref{fig:flx_dis}, and fitted them with a Gaussian function using a $\chi^2$-fit. The reduced $\chi^2$ values for the Gaussian fit to the flux distribution on linear and logarithmic scales are 2.98 and 1.05, respectively, which confirms that the $\gamma-$ray flux distribution of B2 1308$+$326 is log-normal.
A similar log-normal distribution was observed by \cite{2019ApJ...877...39M} for weekly binned $\gamma-$ray light curves of all the six blazars they studied. In Figure \ref{fig:flx_dis}, we also show the distribution of the upper limits as dashed bars.

\subsection{Correlation analysis}
\begin{figure*}
\centering
\includegraphics[width=12cm, height=6.5cm]{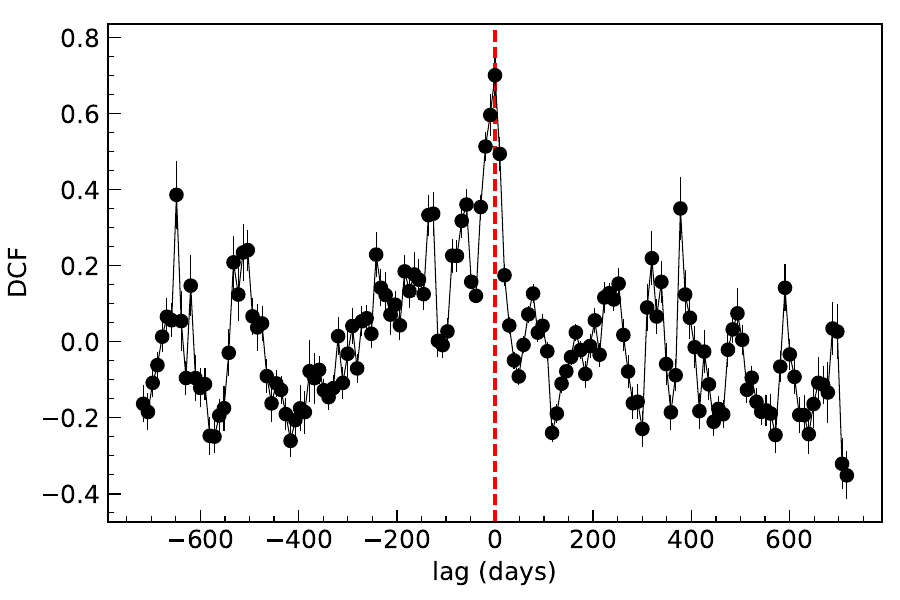}
\caption{\label{fig:dcf}Results of the DCF analysis between the $\gamma-$ray and optical light curves
.}
\end{figure*}

We investigated the correlations between the $\gamma-$ray and optical emissions of B2 1308$+$326 using the discrete correlation function (DCF; \cite{1988ApJ...333..646E}). The DCF method has been extensively used to search for correlations between non-uniform light curves at different wavebands \citep[e.g.][]{2014ApJ...797..137C,2017ApJ...841..123P,2021MNRAS.504.5629R}. The peaks in the DCF plot denote the correlations whose significance increases as the peak value increases towards unity. The result of DCF analysis between the one-day-binned $\gamma-$ray and optical $g-$band light curves is plotted in Figure \ref{fig:dcf}, where a positive lag would indicate that the $\gamma-$rays lead the optical emission. The strong positive DCF peak at about zero time lag indicates that the $\gamma-$ray and optical emissions are strongly correlated. 

\section{Origin of the multi-wavelength emission}\label{sec:sed_modeling}
The multiwavelength data analysed in Section \ref{sec:data} allow us to generate the broadband SEDs of B2 1308$+$326 in different flux states. We generated the SEDs for the following epochs: 
\begin{enumerate}[(a)]
    \item Quiescent state (from MJD 58990 -- 59020): When the $\gamma-$ray flux of the source was below the average flux (5.69 $\times$ 10$^{-8}$ ph cm$^{-2}$ s$^{-1}$) for the entire duration.  
    \item Flaring state (from MJD 59754 -- 59770 ): When the source was in the bright state in all the bands. This period corresponds to a Bayesian block within the flaring duration given by the BB algorithm. 
\end{enumerate}
\begin{figure}
\centering
\includegraphics[width=9cm, height=7cm]{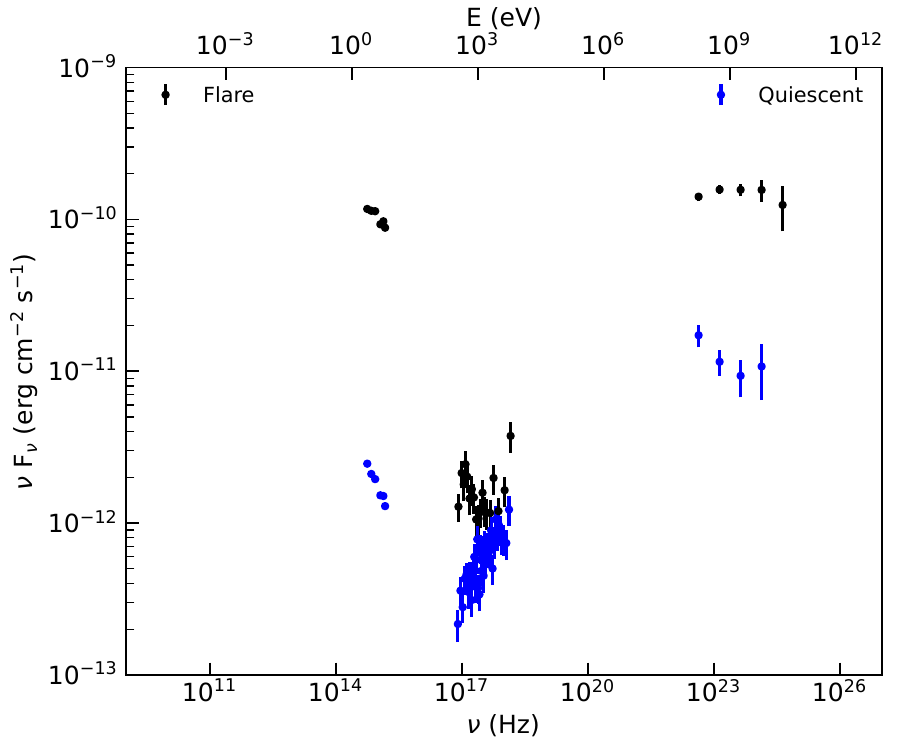} 
\caption{\label{fig:sed_data}Broadband SEDs of B2 1308$+$326 in the quiescent and flaring states specified in the text.}
\end{figure}
The broadband SEDs for these epochs are shown in Figure \ref{fig:sed_data}, which depicts two highly intriguing characteristics. First is a change in the Compton dominance; the $\gamma-$ray peak dominates the synchrotron peak in the low state, while in the flaring state, both peaks are similar. Second is an indication of a shift of the optical-UV synchrotron peak, where the lower state SED is like that of an FSRQ, while in the flaring state, it is closer to that of an ISP.  

To investigate and comprehend the emission mechanism and potential cause of the peak shift, we modelled the broadband SEDs of B2 1308$+$326 using a one-zone leptonic scenario.
In this model, the emission originates from a spherical blob of particles with a radius, $R$, within the jet. The blob is filled with a uniform magnetic field, $B$, and moves down the jet with a bulk Lorentz factor of $\Gamma$ at a small angle $\theta$ to the observer, such that the Doppler factor $\delta \simeq \Gamma$.  The blob is assumed to be filled with relativistic electrons with a broken power-law distribution of the form 
\begin{equation}
    N(\gamma) = \begin{cases}
			N_e \gamma^{-p}, & \gamma_{min} \leq \gamma \leq \gamma_{br}, \\
    N_e \gamma_{br}^{p_1 - p} \gamma^{- p_1}, & \gamma_{br} < \gamma \leq \gamma_{max},
           
		 \end{cases} 
\end{equation}  
where $p$ and $p_1$ are the indices below and above the break energy $\gamma_{\rm br}$.\, $\gamma_{\rm min}$ and $\gamma_{\rm max}$ are the minimum and maximum electron Lorentz factors, respectively, and $N_e$ is the normalization. 

The electrons interact with the magnetic field and produce synchrotron emission, which causes the low-energy component of the SED. The HE part of the SED is generated by the IC scattering of seed photons by the same population of electrons. When the seed photons are the synchrotron photons, the HE component of the SED is explained by the SSC process \citep[e.g.][]{1996ApJ...461..657B}.  On the other hand, if the seed photons come from the external fields such as an accretion disk, BLR, and/or dusty torus (DT) the HE component is caused by EC-disk, EC-BLR, and/or EC-DT, respectively \citep[e.g.][]{1992A&A...256L..27D,1994ApJ...421..153S,2000ApJ...545..107B}. 

The exact location of the $\gamma-$ray emitting region in blazar jets is still unclear. However, we can constrain the location of the emission region, $R_{diss}$, using the observed minimum variability timescale and the energy of the highest-energy photon \citep[e.g.][]{2020ApJ...890..164P}. We detected a minimum variability timescale of $\sim$ 81 hr, which corresponds to a region size of  
\begin{equation}
    R \leq c \tau \frac{\delta}{1+z} \leq 4.38 \times 10^{16} \left(\frac{\delta }{10}\right)\text{ cm}.
\end{equation}
The $R_{diss}$ is then given by $R_{diss}$ =  R/$\psi$, where $\psi$ is the semi-aperture angle of the jet, whose values generally range between 0.1 and 0.25 \citep[e.g.][]{2009ApJ...692...32D,2015MNRAS.448.1060G}. Radio observations have also reported values of $\psi <$  0.1, but radio features do not occupy the whole section of the jet \citep[e.g.][]{2005AJ....130.1418J}. Here, we assumed a typical value of $\psi$ = 0.1 \citep{2015MNRAS.448.1060G}. It is important to note that the BLR is opaque for the high-energy (E > 20 GeV/(1+z)) $\gamma-$rays because they are absorbed via $\gamma\gamma$ pair-production \citep{2006ApJ...653.1089L,2014ApJ...794....8S}. The maximum energy of a $\gamma-$ray photon detected during the flaring state is 27 GeV, which implies that the  $R_{diss}$ location is probably outside the BLR.      

For the broadband SED fitting of B2 1308$+$326, we assumed that the emission region is either at the outer edge of the BLR or outside the BLR and the HE part of the SED is represented by the SSC $+$ EC-BLR $+$ EC-DT. We considered BLR as a spherical shell with inner and outer radii of $R_{BLR}$ and 1.1 $\times$ $R_{BLR}$ \citep{2007ApJ...659..997K}, respectively,  where $R_{BLR}$ = 10$^{17} (L_{\rm disk}/10^{45})^{0.5}$ cm and $L_{disk} = 9 \times 10^{45}$ erg/s \citep{2015MNRAS.448.1060G}. The distance to the dusty torus was assumed to be $R_{DT}$ = 2 $\times$ 10$^{18} (L_{\rm disk}/10^{45})^{0.5}$ cm, and its temperature was fixed at a typical value of 800 K \citep{2015MNRAS.448.1060G}.

\begin{figure*}
\centering
\includegraphics[width=9cm, height=7cm]{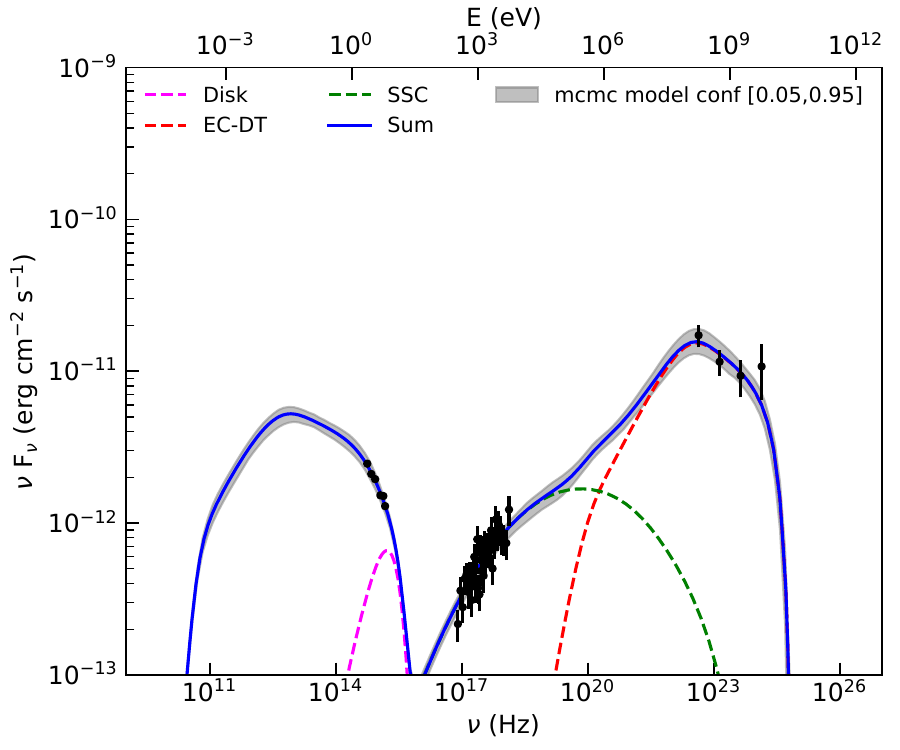} \includegraphics[width=9cm, height=7cm]{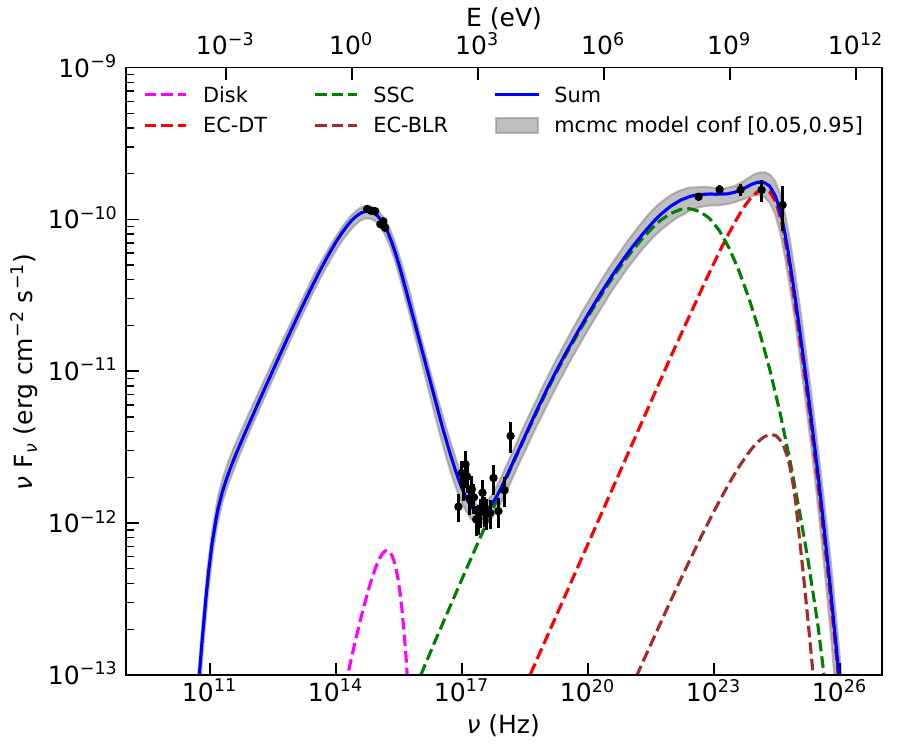}
\caption{\label{fig:seds}Model-fit broadband SEDs of B2 1308$+$326 in the quiescent (left panel) and flaring (right panel) states. The solid blue curve represents the sum of all the components, while the grey shaded region denotes the uncertainty region from MCMC sampling of the model parameters. The legends provide the colour-coding for the other components.}
\end{figure*}

We performed the SED modelling using the publicly available code \jetset \ \citep{2006A&A...448..861M,2009A&A...501..879T,2011ApJ...739...66T,2020ascl.soft09001T}. The free parameters were $p$, $p_1$, $\gamma_{\rm min}$, $\gamma_{\rm max}$,  $\gamma_{\rm br}$, $N_e$, $B$, $R$, and $\delta$, which were constrained during the fitting. The Minuit optimizer was used to limit the parameters initially, and subsequently, MCMC sampling of their distributions was used to enhance them. The values of the best-fitted model parameters for both quiescent and flaring SED modelling are given in Table \ref{tab:sed_res}. The optimal model SEDs for both states are plotted in Figure \ref{fig:seds}.

\begin{table*}
    \centering
    \caption{\label{tab:sed_res}Best-fit model parameters for the broadband SEDs of B2 1308$+$326 in the quiescent and flaring states.}
    \begin{tabular}{lccc} \hline
Parameter                       & Symbol &  Quiescent  & Flare \\ \hline
Low energy spectral index       & p                  		&   1.95$\pm$0.03                & 1.75$\pm$0.04  \\
High energy spectral index      & p$_1$                 	&   3.24$\pm$0.07                & 5.14$\pm$0.16 \\
Minimum electron Lorentz factor & $\gamma_{\rm min}$    	&  49.61$\pm$0.61                & 3.10$\pm$0.23 \\
Maximum electron Lorentz factor & $\gamma_{\rm max} \times 10^4$ & 1.31$\pm$0.04             & 13.42$\pm$1.12 \\
Break energy                    & $\gamma_{\rm br} \times 10^3$	& 1.01$\pm$0.02                  & 6.82$\pm$0.37  \\
Normalization                   & N$_e$  (cm$^{-3}$)        & 19.13$\pm$0.59                 & 239.17$\pm$20.91 \\
Region size                     & R  $\times 10^{17}$ cm  	& 2.32$\pm$0.05                 & 0.75$\pm$0.05 \\
Magnetic field                  & B   (G)                	&   0.14$\pm$0.01                & 0.29$\pm$0.02  \\
Doppler factor                  & $\delta$            		&  15.61$\pm$0.34                & 26.97$\pm$1.14  \\
Region location                 & R$_{diss} \times 10^{17}$ cm  & 10.01                              &  4.05        \\
Electron luminosity             & L$_e$ $\times 10^{45}$ (erg/s) 	           & 3.91          & 4.54   \\
Magnetic field luminosity       & L$_B$ $\times 10^{45}$ (erg/s)               & 0.91	       & 1.25  \\
Proton luminosity               & L$_p$ $\times 10^{45}$ (erg/s)               & 3.53	       & 13.71   \\
Radiation luminosity            & L$_r$ $\times 10^{45}$ (erg/s)               & 0.78	       & 3.52  \\
Total jet luminosity            & L$_{total}$ $\times 10^{45}$ (erg/s)         & 9.13         & 23.02   \\
\hline 
    \end{tabular}
\end{table*}

In the quiescent state, the synchrotron radiation from the jet dominates the optical/UV band, the SSC contributes most in the X-ray band, and the EC-DT produces the $\gamma-$rays (see the left panel of Fig.\ \ref{fig:seds}). The SED modelling shows that the slope of the electron distribution changes from 1.95 to 3.24 at the break energy of 1.01 $\times 10^{3}$. The minimum energy of electrons $\gamma_{min}$ is $\sim$ 49.61, which is a typical value for LSPs \citep{2014ApJS..215....5K}.
The magnetic field is found to be 0.14$\pm$0.01 G. The Doppler factor is 15.61, which is consistent with the value found by \citep{2010MNRAS.402..497G}. The emitting region size is 2.32 $\times 10^{17}$ cm. The synchrotron and IC peak frequencies are $\sim 8 \times$10$^{12}$ and $\sim 5 \times$10$^{22}$ Hz, respectively. The value of the CD is 2.93.

In the flaring state, the synchrotron radiation still dominates the optical/UV band, but the X-ray band receives contributions from both the synchrotron and SSC emissions. The SSC emission extends up to the low-energy Fermi-{\it LAT} band, while the high-energy $\gamma-$ray data are fitted by EC-DT (see the right panel of Fig. \ref{fig:seds}). The power-law slope of the electron distribution at low energy is slightly harder than that in the quiescent state, but the high-energy slope has increased to 5.14 with a higher break energy of 6.82 $\times 10^{3}$. The minimum electron Lorentz factor $\gamma_{min}$ is only $\sim$ 3, indicating that electrons with low energy are also effectively accelerated. 
The strength of the magnetic field is marginally (0.29$\pm$0.02 G) higher than that in the quiescent state, but the Doppler factor has now increased to 26.97. The radius of the emitting region is 7.49 $\times 10^{16}$ cm, which corresponds to a variability timescale of $t_{var} =  R(1+z)/\delta$c $\sim$ 2 days. The peak frequencies for the synchrotron and IC components are $\sim 6 \times$10$^{14}$ and $\sim 2 \times$10$^{24}$ Hz, respectively. The value of the CD is 1.39.

\section{Discussion and conclusions}\label{sec:diss_con}
We extensively investigated the multi-wavelength emission from the transition blazar B2 1308$+$326 with a focus on its most recent outburst in 2022 using data from {\it Fermi}, {\it Swift}, and ASAS-SN. The blazar showed flux variations in each of the studied energy bands. The maximum daily averaged $\gamma-$ray flux of (14.24$\pm$2.36) $\times$ 10$^{-7}$ ph cm$^{-2}$ s$^{-1}$ was detected on MJD 59761.65 during the 2022 flare. The flux variability amplitudes are higher at lower energies. The derived minimum variability timescale in the optical band is $\sim$81 hr (3.4 days), which is shorter than the observed minimum variability timescale of $\sim$141 hr (5.9 days) in the $\gamma-$ray. The inability to identify faster variability in the $\gamma-$ray band may, however, be due to the relatively large error bars. The measured minimum variability timescales are consistent with those observed from the SED modelling.

We observed that the $\gamma-$ray flux of B2 1308$+$326 follows a lognormal distribution. The study of the flux distribution is a unique tool for probing the underlying physical processes. A Gaussian or normal flux distribution, which is expected for linear stochastic variations, suggests additive processes. For non-linear stochastic variations, the flux distribution is log-normal, which more naturally arises from multiplicative processes \citep{2005MNRAS.359..345U}.  Log-normal flux distributions are often found in blazars \citep[e.g.][]{2009A&A...503..797G,2017ApJ...849..138K}. They can be interpreted as the result of multiplicative processes that initially originate in the accretion disk and then propagate to the jet \citep{2008bves.confE..14M}, or as the sum of emission from the randomly oriented mini-jets within the jet \citep{2012A&A...548A.123B}. Variations in the particle acceleration or escape timescales can also produce a log-normal flux distribution \citep{2018MNRAS.480L.116S}. The strong correlation at zero lag found between optical and gamma-ray radiations suggests that they have a common spatial origin.

To understand the origin of the broadband emission of B2 1308$+$326, we generated SEDs in the quiescent and flaring states and fitted them with the one-zone leptonic model. Both SEDs are well represented by the model, defining the low-energy component as synchrotron emission and the high-energy component as the sum of SSC and EC-DT. By comparing the best-fit model parameters for the two SEDs, we found that the electron distribution has a similar low-energy power-law index, but the high-energy indices for quiescent and flaring states are $\sim$ 3.24 and $\sim$5.14, respectively. The differences between the spectral indices, $\Delta p = p_1 - p$ in the two states are 1.29 and 3.39, respectively. The expected value of $\Delta p$ for a standard radiative cooling is 0.5 \citep{1962SvA.....6..317K}. However, \cite{2009ApJ...703..662R} suggested that such a large $\Delta p > 1$ could be naturally produced due to inhomogeneities in the source. The break energy is higher, $\gamma_{br} \sim 6823$, during the flare than in the quiescent state, $\gamma_{br} \sim 1009$, indicating that the particles are accelerated to higher energies during the flare. In the two states, the strength of the magnetic field is comparable, but the Doppler factor has increased from $\sim$ 15.6 to $\sim$27 during the flare. 

We observed that the peak frequencies for the synchrotron and IC components were pushed to higher values during the flare. In the quiescent stage, the synchrotron peak frequency was $\sim 8 \times$ 10$^{12}$ Hz, suggesting that the source was an LSP, but it shifts to $\sim 6 \times$ 10$^{14}$ Hz during the flare, indicating a source transition from LSP to ISP. We also noted a shift in IC peak frequency from $\sim 5 \times$ 10$^{22}$ Hz (quiescent) to $\sim 2 \times$ 10$^{24}$ Hz (flare). A similar shift in the synchrotron peak frequency of B2 1308+326 during a high flux state has also been reported by \cite{2000A&A...364...43W}. This shift in the synchrotron peak frequency during the flare has been observed in several blazars \citep[e.g.][]{2002ApJ...571..226C,2011A&A...529A.145D,2012MNRAS.420.2899G,2014MNRAS.445.4316C}
The transition from one blazar subclass to the next has also been reported using spectroscopic data \citep[e.g.][]{2012ApJ...748...49S,2014ApJ...797...19R,2021AJ....161..196P}. 
The shift from LSP to ISP/HSP can be attributed to (i) variations in the Doppler factor alone \citep[e.g.][]{2009A&A...496..423B}; (ii) hiding of the broad emission lines by overwhelming
synchrotron emission that peaks in the UV in the sources with radiatively weak cooling  \citep{2012MNRAS.425.1371G}; and (iii) swamping of broad emission lines by the variability in the jet continuum emission in the sources with radiatively efficient accretion flows and strongly beamed jets \citep[e.g.][]{2012MNRAS.420.2899G,2014ApJ...797...19R}. The results of our broadband SED modelling suggest that the transition of B2 1308$+$326 from LSP to ISP is due to an increase in the Doppler factor, which also explains the shift in the IC peak and the decrease in the CD.  The change in the Doppler factor can be explained by geometrical effects, such as the change in the viewing angle of the emitting region within the jet \citep[e.g.][]{1999A&A...347...30V,2017Natur.552..374R}. An inhomogeneous and helically curved jet can undergo orientation changes caused by the magnetohydrodynamic instabilities or rotation of the twisted jet. This leads to variations in the viewing angle toward jet-emitting regions, and hence, to variations in the Doppler factor. When the orientation of the emitting region is closely aligned to the observer, the emission from it is more strongly Doppler boosted. 
As a result, a flare can be seen. Using long-term Very Long Baseline Array (VLBA) observations, \cite{2017A&A...602A..29B} found that the jet of B2 1308+326 has a helical structure.
The 2022 flare of  B2 1308$+$326 may therefore be caused by a compact emitting region viewed at smaller viewing angles as compared to the entire jet, which enhances the Doppler factor \citep[e.g][]{2017Natur.552..374R}. 

A similar transitional behaviour was also observed in the blazar PMN J2345$-$1555 by \cite{2013MNRAS.432L..66G}. It is an FSRQ source with a synchrotron component that generally peaks in the far-IR region. However, during the January 2013 flare, the synchrotron peak moved to the optical–UV frequencies, which changed it to a BL Lac source. This transitional behaviour was interpreted as due to the change in the location of the dissipation region, R$_{diss}$, from within the BLR (during the low state) to just outside the BLR (during the flare). The corresponding decrease in the radiative cooling allowed the relativistic particles to reach higher energies, resulting in a shift in the synchrotron and IC component peaks. In our case, the broadband SED modelling suggested that the flare and the shift in the peaks were due to the change in the Doppler factor, which is plausibly caused by the change in the orientation of the emitting region. Although the location of the dissipation region was shifted from 10.01 $\times 10^{17}$ cm (in the low state) to 4.05 $\times 10^{17}$ cm (during a flare), it was outside the BLR in both cases. 

Modelling the SED also provides information about the jet luminosity. The total jet luminosity is defined as \citep{2008MNRAS.385..283C}
\begin{equation}
\begin{split}
    L_{total} & = L_e + L_p + L_B + L_r \\
              & = \pi R^2 \Gamma^2 c (U_e + U_p + U_B + U_r)    
    \end{split},
\end{equation}
where L$_e$ (U$_e$), L$_p$ (U$_p$), L$_B$ (U$_B$), and L$_r$ (U$_r$) are the luminosities (energy densities) of the electron, proton, magnetic field, and radiation, respectively.  On the assumption of one proton per relativistic electron and cold protons, the value of the luminosity for each component and the total jet luminosity are listed in Table \ref{tab:sed_res} for both SEDs. The values of L$_e$/L$_B$ in the quiescent and flaring states are $\sim$ 4.29 and $\sim$3.63, respectively, indicating that the system is close to equipartition. Compared to the quiescent state, the overall jet power is marginally higher when the source is flaring. Additionally, in the flaring state, the overall jet luminosity exceeds the disk luminosity, whereas in the quiescent state, it is comparable to the disk luminosity. \cite{2014Natur.515..376G} found that the total jet power of blazars can be higher than their disk luminosities. Our results concur with their findings.
For a black hole mass of 5.25 $\times$ 10$^8$ M$_{\odot}$ \citep{2014Natur.510..126Z}, the Eddington luminosity for B2 1308$+$326 is L$_{\rm Edd} \simeq$  6.61 $\times$ 10$^{46}$ erg/s. The total jet power in the quiescent state is $\sim$ 14\% of L$_{\rm Edd}$, while it is $\sim$ 35\% of L$_{\rm Edd}$ in the flaring state.

\section{Summary}\label{sec:sum}
We have carried out a multi-wavelength analysis of the high-redshift blazar B2 1308$+$326 for around $\sim$14 yr. Our main findings are summarized as follows.
 \begin{itemize}
     \item B2 1308$+$326 exhibited an historically large flare in optical-to-gamma-ray frequencies in June and July 2022, which reached a maximum daily averaged $\gamma-$ray flux of  1.42 $\times$ 10$^{-6}$ ph cm$^{-2}$ s$^{-1}$ on MJD 59761.65 (1 July 2022).
     \item The estimated fractional variability amplitude was larger at low frequencies (optical/UV) than at high frequencies (X-ray/$\gamma-$ray). The detected variability timescale was also shorter ($\sim$3.4 days) at the optical frequency than the variability timescale ($\sim$6 days) observed at $\gamma-$rays.
     \item The $\gamma-$ray flux distribution of the source followed a log-normal distribution.
     \item The optical and $\gamma-$ray emissions are positively correlated without any detectable time lag.
    
    \item The synchrotron peak frequency increased from  $\sim 8 \times$10$^{12}$ Hz (in the quiescent state) to $\sim 6 \times$10$^{14}$ Hz (in the flaring state) with a corresponding decrease in the CD indicating that the source changes from LSP to ISP during the flare. 
    \item The SED modelling suggested that the transition was most likely produced by an increase in the Doppler factor.
 \end{itemize}

\begin{acknowledgements}
We thank the anonymous referee for their valuable comments and suggestions which helped to make the manuscript better. Part of this work was supported by the Polish Funding Agency National Science Centre, project 2017/26/A/ST9/00756 (MAESTRO 9). This project has received funding from the European Research Council (ERC) under the European Union’s Horizon 2020 research and innovation program (grant agreement No. [951549]).
\end{acknowledgements}


\bibliographystyle{aa}
\bibliography{master} 
\end{document}